# Low-Temperature Rapid Synthesis and Superconductivity of Fe-Based Oxypnictide Superconductors

Ai-Hua Fang,[†] Fu-Qiang Huang,*[,†] Xiao-Ming Xie,[‡] and Mian-Heng Jiang[‡]

*CAS Key Laboratory of Materials for Energy Conversion, Shanghai Institute of Ceramics, Chinese Academy of Sciences, Shanghai 200050, P.R. China, and State Key Laboratory of Functional Materials for Informatics, Shanghai Institute of Microsystem and Information Technology, Chinese Academy of Sciences, Shanghai 200050, P.R. China*

Received January 4, 2010; E-mail: huangfq@mail.sic.ac.cn

The discovery of Fe-based high $T_c$ superconductor by Hosono et al[1] has stimulated extensive research in this field. However, the synthesis of the subject oxypnictide $Ln$FeAsO ($Ln$ = rare earth elements) usually requires a very high temperature and a very long time. For example, when La$_2$O$_3$, La, and FeAs were used as the starting materials, the solid-state reactions took place at 1523 K for 40 h, and there were still impurity phases including FeAs remaining in the resultant LaFeAsO samples.[1] Furthermore, under these synthesis conditions, the preparation of F-doped LnFeAsO samples, which may raise $T_c$, suffered from poor compositional control due to the high volatility of F-containing compounds and cross contamination from F-caused corrosion products in the reaction vessel.[2] Since long reaction time at high temperature and poor compositional control may inadvertently affect the superconductivity and reproducibility of the samples, a method that enables more rapid synthesis at lower temperatures will be of great interest.

The difficulty of $Ln$FeAsO synthesis is probably rooted in the large disparity of the electronegativity of its constituents. Generally, a highly electropositive element (e.g., $Ln$) tends to form a stable ionic compound with a highly electronegative element (e.g., O), whereas a slightly electropositive element (e.g., Fe) tends to form a stable covalent compound with a slightly electronegative element (e.g., As.) Subsequent reaction between the two stable phases, $Ln_2$O$_3$ and FeAs, is usually sluggish, but this difficulty can be largely removed if less stable $Ln$As and FeO are used as the starting materials from which a displacement reaction can take place to form the requisite $Ln_2$O$_2$−Fe$_2$As$_2$ layered compound. The kinetics of reaction can be further accelerated by ball-milling of the raw materials, which can drastically increase the surface area and therefore the reactivity of the reagents. With these considerations, we have successfully synthesized pure-phase Fe-based oxypnictide superconductors at temperatures as low as 1173 K at times as short as 20 min. Such a method of reactive sintering is very attractive for the large-scale economic production of this family of superconductors.

Sample preparation is shown in ref 3. Powder X-ray diffraction (XRD) patterns of SmFeAsO samples prepared under different conditions are shown in Figure 1a. A reference sample with Sm$_2$O$_3$, Sm, and FeAs as the starting materials was fabricated at 1433 K for 20 min for comparison; it contains Sm$_2$O$_3$ and FeAs impurities, and its amount of SmFeAsO is small (curve **A** in Figure 1a). These impurity phases still remained after 40 h sintering at 1433 K. In contrast, when SmAs and FeO were used as the starting materials, SmFeAsO was already the main phase of the sample sintered at 1173 K for only 20 min (curve **B**). The reaction can be further enhanced by mechanical ball-milling of the starting materials (SmAs and FeO), as evidenced by the pure phase of SmFeAsO in the sample obtained after 20 min sintering at 1173 K using powders milled for 2 h (curve **C**). The method is also applicable for F-doped samples; two XRD results of the samples are shown as curves **D** and **E** in Figure 1a, and the $T_c$ measurement of several samples thus synthesized are shown in Figure 1b. Previously, McCallum et al. also used LaAs, Fe$_2$O$_3$, and Fe as starting materials to synthesize $Ln$FeAsO.[6] However, they used a slower heating rate which allowed intermediate phases to form, hence, dissipating the driving force. In contrast, the method of direct reaction at high temperature in our procedure produced much superior results. A group of samples were made by this method with various compositions, $Ln_{1−x}Ln'_x$FeAsO$_{1−x}$F$_x$, $Ln$FeAsO$_{1−x}$F$_x$, and different sintering conditions; the results of $T_c$ measurements of these samples are summarized in Table 1. The $T_c$ referred to henceforth is the onset temperature $T_c^{onset}$ determined by the intersection of the two tangents on the temperature-dependent resistivity curve as illustrated in Figure 1b.

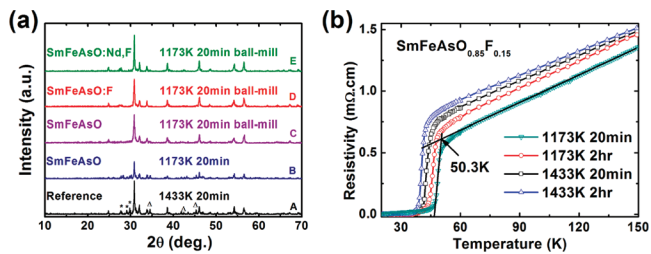

***Figure 1.*** (a) Powder X-ray diffraction patterns of SmFeAsO and F-doped SmFeAsO reactive-sintered under different conditions. Sm$_2$O$_3$ and FeAs impurities are marked by * and Δ, respectively. (b) Temperature-dependent resistivity of SmFeAsO$_{0.85}$F$_{0.15}$ sintered under different conditions from 2 h-ball-milled powders. The straight lines tangent to the bottom curve illustrate how $T_c^{onset}$ is determined.

As we can see from Table 1 samples of the same composition that are prepared under different conditions can have very different $T_c$'s. As mixed lanthanide doping is rarely studied in the literature, we take Sm$_{0.85}$Nd$_{0.15}$FeAsO$_{0.85}$F$_{0.15}$ as the main composition of the sample to investigate the process-related correlation of $T_c$. For samples without ball milling synthesized at the same temperature of 1433 K, $T_c$ first increased with increasing sintering time (from 31.8 K at 20 min to 38.9 K at 2 h) and then decreased at prolonged sintering time (down to 31.9 K at 4 h). This can be explained by the fact that F is gradually incorporated into the lattice as the solid-state reaction proceeds. While the F concentration may reach a maximum value at a certain point, F escapes from the sample at prolonged sintering time, and the F concentration decreases as does $T_c$.

---

[†] Shanghai Institute of Ceramics.
[‡] Shanghai Institute of Microsystem and Information Technology.

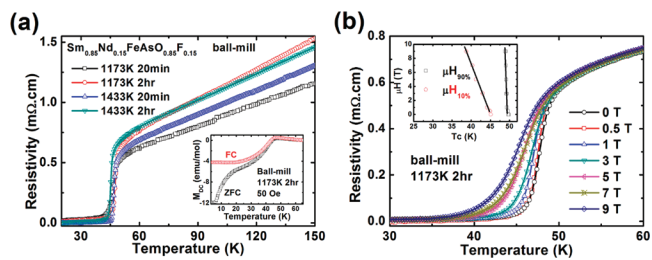

**Figure 2.** (a) Temperature-dependent resistivity of $Sm_{0.85}Nd_{0.15}FeAsO_{0.85}F_{0.15}$ sintered from ball-milled powders under different conditions. (Inset) Magnetic measurements (50 Oe) of $Sm_{0.85}Nd_{0.15}FeAsO_{0.85}F_{0.15}$ sintered at 1173 K for 2 h. (b) Magnetic field dependence of resistivity of $Sm_{0.85}Nd_{0.15}FeAsO_{0.85}F_{0.15}$ sintered at 1173 K for 2 h. (Inset) $\mu H_{c2}$ vs temperature of the same sample determined at 90% and 10% drop of the resistivity.

**Table 1.** Reactive-Sintering Temperature ($T$), Time ($t$) and Onset Temperature of Superconductivity ($T_c^{onset}$) of SmFeAsO-Based Superconductors

| compound | $T$ (K) | $t$ (h) | $T_c^{onset}$ (K) |
|---|---|---|---|
| $Sm_{0.85}Nd_{0.15}FeAsO_{0.85}F_{0.15}$ (no ball-milling) | 1433 | 0.33 | 31.8 |
| | 1433 | 0.5 | 33.2 |
| | 1433 | 1.0 | 37.7 |
| | 1433 | 2.0 | 38.9 |
| | 1433 | 4.0 | 31.9 |
| $Sm_{0.85}Nd_{0.15}FeAsO_{0.85}F_{0.15}$ (ball-milled powders) | 1433 | 0.33 | 48.3 |
| | 1433 | 2.0 | 46.7 |
| | 1173 | 0.33 | 50.7 |
| | 1173 | 2.0 | 50.2 |
| $SmFeAsO_{0.85}F_{0.15}$ (ball-milled powders) | 1433 | 0.33 | 45.5 |
| | 1433 | 2.0 | 42.5 |
| | 1173 | 0.33 | 50.3 |
| | 1173 | 2.0 | 48.2 |
| $SmFeAsO_{0.85}F_{0.15}$ (ref 4) | 1433 | 40.0 | 43 |
| $SmFeAsO_{0.85}F_{0.15}$ (6 GPa, ref 5) | 1523 | 2.0 | 55 |

The advantage of ball-milling is to enhance the reaction kinetics. Therefore, as you can see from Table 1, all $T_c$ values for samples synthesized from ball-milled powders are higher than those without ball-milling. The highest $T_c$ of 50.7 K was obtained at an even lower temperature of 1173 K for only 20 min sintering. The similar mechanism of F loss explained previously can also be seen from the $T_c$ changes of this group of samples: (i) at the same temperature, $T_c$ decreases as sintering time increases (48.3 to 46.7 K at 1433 K and 50.7 to 50.2 K at 1173 K from 20 min to 2 h sintering); (ii) for the same sintering time, $T_c$ drops with increasing temperature (50.7 to 48.3 K for 2 h sintering and 50.2 to 46.6 K for 20 min sintering from 1173 to 1433 K); (iii) the decrease of $T_c$ over sintering time at a lower temperature of 1173 K (0.5 K from 20 min to 2 h) is smaller than that at 1433 K (1.6 K from 20 min to 2 h).

Temperature-dependent resistivity of some samples is shown in Figure 2a. The calculated Meissner fraction of the 1173K-2h sample in the inset is about 2.5%, assuming a theoretical density of 7.1 g·cm$^{-3}$. As shown in Figure 2b, the transition width broadened from about 4.6 K at $0T$ to 10.2 K at $9T$. The characteristic critical fields were determined for $H_{90\%}(T)$ and $H_{10\%}(T)$, the field when the respective resistivities drop to 90% and 10% of the normal-state value. The values of $\mu_0 dH_{90\%}/dT$ and $\mu_0 dH_{10\%}/dT$ of the 1173 K sample are 11.2T/K and 1.4T/K, respectively, as shown in the inset of Figure 2b. Using the Werthamer–Helfand–Hohenberg formula[7] $\mu_0 H_{c2}(0) = -0.69 T_c (d\mu_0 H_{c2}/dT)_{T=T_c}$, we determined the upper critical field to be 392$T$, higher than the 1433K-2h sample (242$T$) and comparable to values found in other Fe-based superconductors.[8,9] The low Meissner fraction and rather high upper critical field can be well-explained by small grain size and poorly crystallized grains in the samples, as shown in Figure S2 of the Supporting Information.

As shown in the last group of data in Table 1, we compare our samples of $SmFeAsO_{0.85}F_{0.15}$ prepared by ball-milled powders sintered at lower temperatures for shorter times ($T_c$ = 50.3 K for 1173K-20min sintering) with those reported either by the conventional solid-state reaction method ($T_c$ = 43 K for 1433K-40h sintering)[3] or by the high-pressure process ($T_c$ = 55 K for 1523K-2h-6-GPa sintering).[4] While the highest reported $T_c$ of 55 K thus far was obtained by the high-pressure and high-temperature sintering process, the advantage of ball-milled powders with lower temperature and shorter time of sintering process leaves room for further improvement of superconducting Fe-based oxypnictide materials. By increasing the F content, we raised the $T_c$ to 53.2 K in the $SmFeAsO_{0.8}F_{0.2}$ sample. Details are in Table S2 and Figure S3 of the Supporting Information.

In conclusion, we were able to develop a novel method to synthesize Fe-based oxypnictide superconductors. By using *Ln*As and FeO as the starting materials and a ball-milling process prior to solid-state sintering, $T_c$ as high as 50.7 K was obtained with the sample of $Sm_{0.85}Nd_{0.15}FeAsO_{0.85}F_{0.15}$ prepared by sintering at temperatures as low as 1173 K for times as short as 20 min. We believe this novel method will enable us to explore the new Fe-based oxypnictide materials for making further improvements to the superconducting properties.

**Acknowledgment.** Financial support from Knowledge Innovation Program of the Chinese Academy of Sciences (Physical Properties and Mechanism of Iron-Based Superconductors) is gratefully acknowledged.

**Supporting Information Available:** Methods for sample preparation and Rietveld refinements, property measurements, resistivity–temperature measurements of selected samples at different magnetic fields and tabulation of $T_c^{onset}$. This material is available free of charge via the Internet at http://pubs.acs.org.